\documentclass[aps,pra,twocolumn,groupedaddress]{revtex4}
\usepackage{graphicx}

\usepackage{latexsym}
\begin{document}

\title{Light scattering from a magnetically tunable dense random medium with weak dissipation : ferrofluid }
\author{M.Shalini}
\affiliation{Centre for Excellence in Basic Sciences, University of Mumbai, Santa Cruz East, Mumbai 400 098, India}
\author{Avinash A. Deshpande}
\affiliation{Raman Research Institute, Sadashiva nagar, Bangalore-560080}
\author {Divya Sharma}
\affiliation{BITS, Pilani-Hyderabad Campus, Jawahar Nagar, Hyderabad 500 078, India}
\author{D. Mathur}
\affiliation{Tata Institute of Fundamental Research, Homi Bhabha Road, Mumbai 400 005,
India}
\author{Hema Ramachandran, and N.Kumar}
\affiliation{Raman Research Institute, Sadashivanagar, Bangalore-560080}


\date{\today}

\begin{abstract}
We present a semi-phenomenological treatment of light  transmission
 through and its reflection from a ferrofluid, which we regard as a magnetically
 tunable system of dense random dielectric scatterers with weak
 dissipation. Partial spatial ordering 
 is introduced  by the application  of  a
 transverse  magnetic field that superimposes a periodic modulation  on the
 dielectric  randomess. This  
 introduces  Bragg
 scattering which effectively enhances  the scattering due to
 disorder alone,  and thus reduces the elastic mean free path towards Anderson localization. Our theoretical
 treatment, based on invariant imbedding, gives a  simultaneous decrease of 
 transmission and reflection without change of incident linear polarisation as the spatial order is tuned magnetically to
 the  Bragg condition, namely the  light wave vector being equal to half the Bragg
 vector (Q).  Our
  experimental observations are in  qualitative agreement
 with these results. We have also given expressions for the transit (sojourn) time of light and for the light energy stored in the random medium under steady illumination. The ferrofluid thus provides an interesting physical realization of effectively a ``Lossy Anderson-Bragg'' (LAB) cavity with which to study the effect of the interplay of spatial disorder, partial order and weak dissipation on light transport.  Given the current interest in propagation, optical limiting and
storage of light in ferrofluids, the present work seems topical.
\end{abstract}
\maketitle

\section{Introduction}
Ferrofluids (magnetic  nanoparticles  dispersed in  liquids) 
display very unusual magneto-optical properties (optical limiting, switching, and
such like) that can be tuned by varying an externally applied magnetic field. In
this work we report on first results of a combined experimental and theoretical
study of light transmission through and its back-reflection from a magnetically
tunable ferrofluid - a random suspension of  magnetite $(Fe_3O_4)$
nanoparticles in kerosene. This  remarkable  system permits   spatial rearrangement of the particles  magneto-statically so as to alter, via the  structure factor, the effective scattering of light in the system. The dual role of the nanoparticles, acting both as dielectric Rayleigh scatterers and as permanent magnetic dipoles,  is a salient
feature of the  nanofluidic optical system, inasmuch as it  scatters 
dielectrically and  may be ordered  magneto-statically.\\ 
 The phenomenon of coherent multiple scattering of waves, like  de Broglie (dB)
electron waves or light waves in a disordered medium, has been extensively studied in the context
of Anderson localization \cite{Anderson, Sheng}. Here, in the limit of strong scattering, the elastic
transport mean free path ($\ell_e$) can decrease to the extent that the wave becomes
spatially localized. This is the well-known  Ioffe-Regel limit, $k_0\ell_e\sim$1,
where $k_0$ (=2$\pi$/$\lambda_0$) is the magnitude of the wave-vector corresponding
to the wavelength $\lambda_0$ in the medium. In the case of electrons, 
strong scattering ($\ell_e < \lambda_{dB}$) results in localization
of the electron waves \cite{Graham}, as in the metal-insulator transition. For light, however,
such strong scattering demands an unrealistically high dielectric (refractive index)
mismatch, making it difficult to localize light \cite{JohnPhysToday}.  At very long
wavelengths, localization escapes the Ioffe-Regel condition because of the weakness of
Rayleigh scattering ($\lambda_0^{-4}$). At very short wavelengths geometrical
(ray) optics holds and there is no scope for localization, which is essentially a
multiple wave-interference effect.  The localization condition for light may, however, 
be closely approached  by introducing a periodic modulation in the disordered medium
 such that the Bragg scattering opens a forbidden optical band-gap, presumably  a
pseudogap,  giving  a small $\ell_e\sim \lambda_0$ due to enhanced random
scattering. Indeed, this has been shown for  photonic bandgap materials in the
presence of disorder \cite{JohnPRL}. In our experiments, we combine disorder (dense random
scattering) with partial order (Bragg scattering) in the presence of weak
dissipation obtaining  in a ferrofluid placed in a magnetic field. We induce spatial
order that can be magneto-statically tuned towards Bragg resonance so as to give
rise to a much smaller $\ell_e$,  causing  the time spent by light within the medium  to be significantly enhanced. In the presence of weak dissipation (inevitably present
in a ferrofluid), we demonstrate, experimentally as well as theoretically, the simultaneous
decrease of transmission and reflection of light over a range of disorder
($\ell_e$-values). One may aptly refer to our  ferrofluidic sample, essentially a weakly dissipative random suspension of $n_{FF}$ nanomagnets per unit volume, with a weak spatial periodicity superimposed, as a Lossy Anderson-Bragg (LAB)
cavity. Our dielectric nanomagnets  scatter
light by virtue of their complex (lossy) dielectric polarisability, presenting us
with a model system that has in it an interplay of spatial disorder, partial order, and
weak dissipation.\\
\indent In this work, we first present a brief discussion of the phenomenology of the
long photon transit (sojourn) time that we expect in a ferrofluid due to multiple
scattering. This is followed by 1D numerical simulation of light transport
 based on invariant imbedding. We then present  experimental results for our
tunable ferrofluidic LAB cavity, which are found to be in general agreement with  our theory.\\ 
\section{Phenomenology}

\begin{figure*}
\begin{center}
 \includegraphics[width=16cm,height=7cm]
{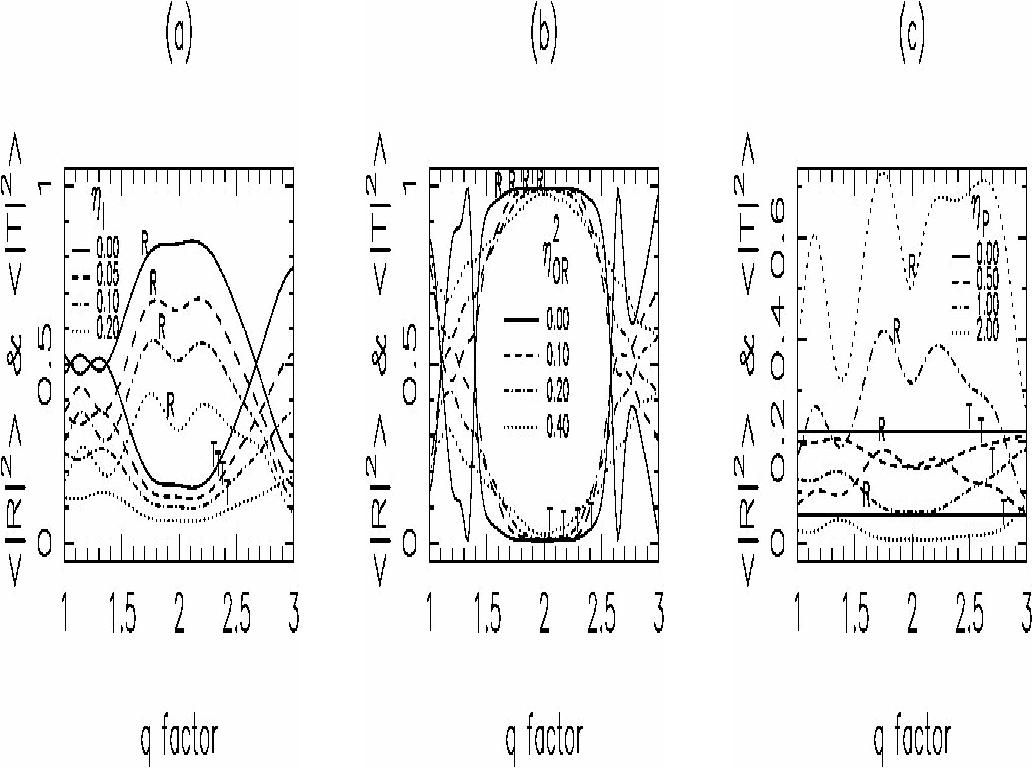}
\caption{ (a) Plots of  transmission $|T|^2$ and  reflection $|R|^2$ coefficients
(using eqs. \ref{e.dRbydx}, \ref{e.dTbydx}) as function of the tuning parameter $q$ for   different
values of the dissipation parameter ($\eta_I$). Note the dips in both $|T|^2$ and
$|R|^2$ at $q = 2$, corresponding to maximum scattering (Bragg condition). Here the disorder parameter
$\eta^2_{OR} =0.15$, the depth of modulation $\eta_P =1$, and the dimensionless
sample length $x=7$;  (b) Plots of  transmission $|T|^2$ and  reflection $|R|^2$ coefficients
(using eqs. \ref{e.dRbydx}, \ref{e.dTbydx}) as function of the tuning parameter $q$
for  different values of the disorder parameter ($\eta^2_{OR}$). Note the optical stop band bracketting $q=2$. Here the dissipation $\eta_I = 0$, the
depth of modulation $\eta_P =1$, and the dimensionless sample length $x=14$; (c) Plots of  transmission $|T|^2$ and  reflection $|R|^2$ coefficients
(using eqs. \ref{e.dRbydx}, \ref{e.dTbydx}) as function of the tuning parameter $q$ for different depths of periodic modulation, $\eta_P$. Here the dissipation parameter $\eta_I = 0.2$, the disorder parameter $\eta_{OR} = 0.15$ and the dimensionless sample length $x=7$. The dominant role of the periodic modulation is clearly brought out here.} 
\end{center}
\label{f.desh1}
\end{figure*}
 Considering a LAB cavity of linear dimension $ \sim L$, we examine the transmission through and reflection from it. The mean time $\tau_L$ for the
photon to diffusively traverse the LAB cavity is given by  2$D\tau_L$=$L^2$,
with the diffusion constant, $D= c_0\ell_e$, $c_0$ being the speed of light in
the medium.
The actual path length traversed by the photon is then $\Lambda=\tau_Lc_0$. With inelastic scattering mean free path $\ell_i$,
the survival probability, $p_s$,  for a photon injected into our LAB cavity becomes
$p_s= e^{-\Lambda/\ell_i} = e^{-L^2/(2\ell_e\ell_i)}.$
It is this surviving photon that finally 
emerges from the sample --  in transmission ($T$) or in reflection ($R$) -- for a 1D
system. For many multiple scatterings, an  equipartition between the emergent
quantities  $R$ and $T$ is expected, and consequently, 
\begin{equation}
\label{e.ReqT}
R\sim e^{-L^2/(2\ell_e\ell_i)}  \sim T
\end{equation}
Clearly, $R$ and $T$  decrease as the
ferrofluid is tuned towards stronger scattering. We note that, strictly speaking,  Eq. 1 is valid for a system where the forward and the 
backward directions are clearly defined. Further, it pertains to the case where
photons are injected well within the optical sample whose linear dimensions are much
greater than $\ell_e$. In a real experiment, of course, as also in the simulation to follow, photons are incident at one end of the sample. This leads to an asymmetry in that typically the reflected photons traverse smaller path lengths in the LAB cavity than the transmitted ones; thus  $R$ diminishes to a smaller extent than
$T$ due to the dissipative effects.  

The above  phenomenology remains  essentially valid  for  a 3D  system of  \textit{dilute } ($
{{4 \pi}\over3} \lambda_0^3 n_{FF} < 1$) random  nanoparticulate (NP) Rayleigh  
scatterers (radius $a_{NP}  \ll \lambda_0$). For such a system, the elastic and the inelastic mean free paths 
can be readily expressed in terms of the real and the imaginary parts of the optical
dielectric constant ($\epsilon  = \epsilon_R + i  \epsilon_I$ ) of the dielectric
nanospheres in the ferrofluid.  Explicitly, we have, 
$ \ell^{-1}_e  = n_{FF} \sigma,  $
with  the scattering cross-section, $\sigma$, of the nanoparticles  being given by 
\begin{equation}
\label{e.sigma}
 \sigma  = {\frac{8 \pi}{3}} k_0^4 a_{NP}^6  {\frac{( \epsilon_R - 1)^2}{
(\epsilon_R +2)^2}}
\end{equation}
and, 
$ \ell^{-1}_i =  (2 \pi \epsilon_I) / (\lambda_0 \epsilon_R )$

In the case of our ferrofluid,  however, we have the opposite limit of a \textit{dense}
random scattering system ($ {{4 \pi} \over3 } \lambda_0^3 n_{FF}  > 1$) with the number
density $n_{FF}$ of the scatterers $\sim 10^{16}cm^{-3}$ and wavelength of light,
$\lambda$, $633 nm$. In such a case, the scatterers
lying within a wavelength  scatter essentially in the forward direction, and  coherently, calling for a
different treatment as presented below for a 1D model system. 
\section{1D simulation  based on invariant imbedding }
We simulate our 1D model system  using invariant imbedding \cite{Kumar, Ramal}
giving directly the emergent quantities $R$ and $T$ as functions of the sample
length. Here we introduce a position ($\ell$)-dependent complex dielectric constant,
$\epsilon(\ell)$ that comprises three parts: a disordered part ($\epsilon_R(\ell)$)
which is random and real, accounting for elastic scattering; an ordered part,
($\epsilon_P(\ell)$) which is periodic  and real, representing  Bragg scattering; and
a dissipative imaginary part, $i\epsilon_I(\ell)$ quantifying  the loss. 
Also, $\epsilon_0$ represents the real part  of  the medium's mean (background)
dielectric constant,  giving the wave velocity $c_0=c/\sqrt{\epsilon_0}$, and the
wavevector $k_0=k\sqrt{\epsilon_0}$=2$\pi/\lambda_0$. As is appropriate for a
dielectric at optical frequencies, the magnetic permeability has been set to unity.  Thus, we have
for our 1D model ferrofluid  
\begin{equation}
\label{e.epsilon}
\epsilon(\ell)=\epsilon_0+i\epsilon_I + \epsilon_R(\ell) + \epsilon_P, 
\end{equation}

{\noindent}with 0$<\ell<$L. We model $\epsilon_R(\ell)$ as a
$\delta$-correlated random Gaussian variable, and $\epsilon_P$ as a sinusoid of
wavelength $2\pi / Q$. The emergent quantities, $R$ and $T$, can now be obtained as
function of the sample length, $L$, by the method of invariant imbedding 
in the following equations that incorporate all the essential features of our model:
the real random disorder ($\eta_{R}$), the ordered periodic modulation ($\eta_P$),
and the dissipation ($\eta_I$) : 
\begin{equation}
\label{e.dRbydx}
\frac{dR(x)}{dx}= 2iR(x) + \frac{i}{2}[1+R(x)^2][\eta_{R}(x)+ 
\eta_P
sin(qx)+i\eta_I]
\end{equation}
and
\begin{equation}
\label{e.dTbydx}
\frac{dT(x)}{dx} =
 iT(x) +
\frac{i}{2}[1+R(x)T(x)][\eta_{R}(x)+\eta_P sin(qx)+i\eta_I],
\end{equation}
where the dimensionless variables and normalized parameters are introduced as 
$x=k_0L,~q=Q/k_0,~\eta_{R}(x)=\epsilon_R(L)/\epsilon_0,~\eta_P=\epsilon_P/\epsilon_0,
~and~\eta_I=\epsilon_I/\epsilon_0$.

For the random variable $\eta_{R}(x)$, we have $<\eta_R(x)>$=0,
$<\eta_R(x)\eta_R(x')>=\eta_{OR}^2\delta(x-x')$. Results of our numerical
solution of the stochastic eqs. \ref{e.dRbydx} and \ref{e.dTbydx} are shown in 
Fig. 1 in the form of
transmission ($\left|T\right|^2$) and reflection ($\left|R\right|^2$) coefficients
as function of the tuning parameter $q$ for fixed sample length. This is shown in Fig. 1(a) for different values of the dissipation parameter $\eta_I$, for fixed disorder and fixed depth of periodic spatial modulation. 
Fig. 1(b) shows the corresponding dependence on $q$ for
different values of disorder, for a fixed depth of  modulation  and zero dissipation,  and 
Fig.1(c) for different depths of periodic modulation.  (A technical
point to be noted here is that in our computation we have used the polar form for 
the complex quantities $R$ and  $T$, so as to simultaneously monitor both the magnitudes and
the phases separately. This allows us to ensure the physical upper bound of unity for both
$\left|R\right|^2$ and $\left|T\right|^2$, for all realizations of disorder,
dissipation  and for all lengths). 
Note in Fig. 1(a) that when dissipation is not too small,
both $\left|R\right|^2$ and $\left|T\right|^2$ dip as the Bragg condition ($q=2$) is
approached,  with the dip being more pronounced for the former. Note also that
$\left|T\right|^2$ remains smaller than $\left|R\right|^2$, with the sum
remaining less than unity (non-zero dissipation). In 
Fig. 1(b), the opening of the
pseudo-bandgap is clearly discernible ($\left|R\right|^2$ is nearly unity and
$\left|T\right|^2$ is nearly zero), with the sum remaining unity (zero dissipation).
 In 
Fig. 1(c) it is seen that periodic modulation plays a dominant role in the scattering.\\
It is now instructive to compare these results of our 1D numerical simulation  with
those from our experiments.  Here we emphasize that the 1D system is
physically equivalent to a 3D system, with the proviso that all spatial
fluctuations and modulations be taken along the direction of incidence of the light.
This is indeed approximately the case for our ferrofluidic system inasmuch as the
transverse magnetic field is expected to induce spatial ordering of the interacting
nanomagnets, while the dense random dielectric scatterers scatter predominantly in the forward direction. Indeed, the
Bragg scattering is the dominant determining feature as far as multiple scattering
is concerned.

\section{Experiments}
\begin{figure}
 \centering
 \includegraphics [width=8cm,height=6cm]
 {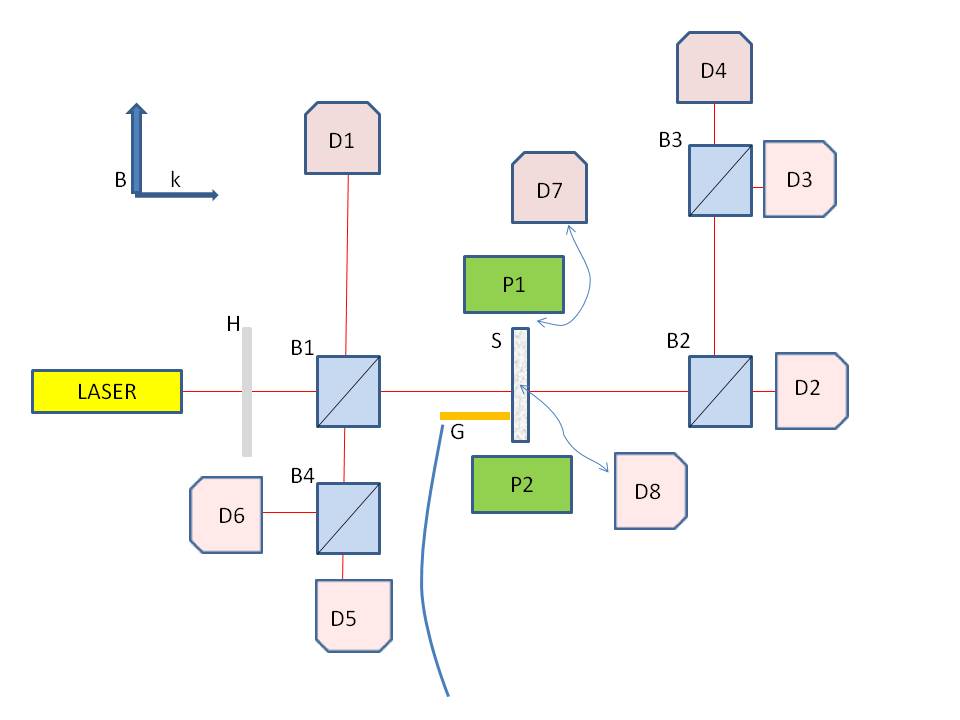}
 \caption{Schematic  of the experimental setup, showing sample (S) kept between the pole pieces P1 and P2 of an electromagnet. G is the Gauss Probe, H the half-wave plate, D1-D8 detectors, and B1 and B2 are non-polarising beam splitters while  B3 and B4 are polarising beam splitters.  Arrows {\bf{B}} and {\bf{k}} denote the directions of the applied magnetic field and the wave-vector of the incident light. }
 \label{f.schematic}
\end{figure}

We have measured the reflected and the transmitted intensities and their polarisations
upon irradiation of our ferrofluid by a $633 nm$ light from a $5 mW$ cw He-Ne laser.
The ferrofluid (Ferrolabs, USA) consisted of surfactant coated $Fe_3O_4$ nanospheres of radius $a_{NP}$ ($ \sim 10nm$) randomly dispersed in kerosene.  Figure \ref{f.schematic}  is the
schematic  of our apparatus. Linearly polarised light from the laser source (the plane of polarisation of which could be rotated by means of the halfwave plate, H) was incident on the sample (S) that was contained in a 
 glass cuvette of size $1cm \times 1cm  \times  2mm$ ($2mm$ along the direction of incident light).  A non-polarising beam-splitter (B1) placed in the path of the incidence allowed  us to  monitor the intensity of the incident light using detector D1 and that of the reflected light using detectors D5 and D6. The cuvette was  held between the pole pieces 
(P1 and P2) of
an electromagnet that could produce a tunable magnetic field tranverse to the direction of
incidence of light, of upto ~450G, uniform over the illuminated sample to within a
few mG, as measured by the Hall-effect Gauss probe (G). {The light transmitted through the sample was incident on a non-polarising beam-spiltter (B2); detector D2 provided a  measure of the transmitted intensity while the combination of the polarising beam-splitter (B3) and detectors D3 and D4 enabled the determination of its plane of polarisation. Similarly, using polarising beam splitter B4 and detectors D5 and D6, the intensity and polarisation of the reflected light could be determined.  Detectors D1 - D6 were identical photodetectors (Thorlabs DET110).  A small photodiode (D7) was placed in the narrow gap between the electromagnet and the cuvette in order to measure the light intensity scattered to the side. A similar detector (D8) placed on top (perpendicular to both the direction of light incident on the sample and to the applied magnetic field) was used to measure the light intensity scattered in that direction.

\begin{figure}
 \centering
 \includegraphics [width=8cm,height=6cm 
] {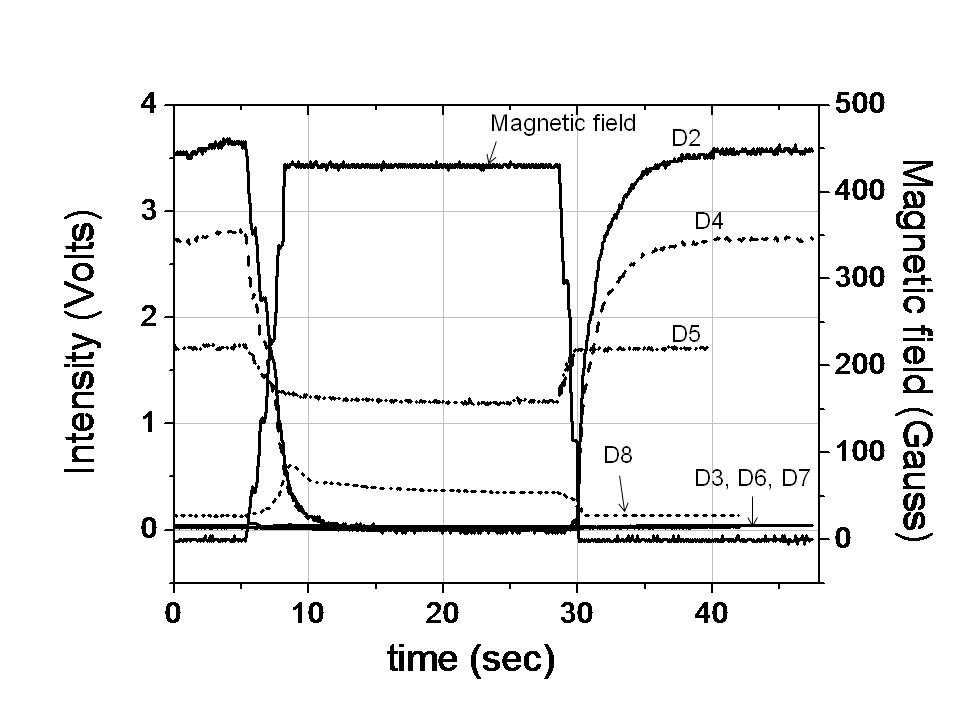}
 \caption{Oscilloscope traces showing, as function of time, the applied magnetic field and the light intensities as measured by detectors D2-D8. The inensity of the transmitted light  reduces to zero when the magnetic field is increased; that of the  reflected light is found to decrease to a smaller extent.  Both the transmitted and the reflected light retain the  incident linear polarisation, in this case horizontal polarisation.}
 \label{fig:2}
\end{figure}

{Figure 3 shows the relative variation of the reflected and the  transmitted
intensities and their polarisations with applied magnetic field.  In this case the incident light is horizontally polarised (\textit{i.e.}, plane of polarisation perpendicular to the incident wave-vector, k, and parallel to the applied magnetic field, B).  A simultaneous
decrease in both the transmitted light (D2) and the reflected  light (D5) is observed as the applied field is increased. 
While the transmission drops to near-zero, the reduction in the reflection is less
pronounced. On decreasing the magnetic field back to zero, the original intensities are
recovered.  The intensities recorded on detectors D4 and D5 (horizontal polarisation) and D3 and D6 (vertical polarisation) confirm that the transmitted and the reflected light retain the input (horizontal) polarisation. The scattered intensities in the transverse
directions (D7, D8) are quite small. The situation was quite similar when the plane of polarisation of the incident light was perpendicular to the applied magnetic field (\textit{i.e}., vertical).\\ 
The experiment was then repeated for two
other wavelengths - (a) $612 nm$ (He-Ne laser) for which the fall of the intensities with
increasing magnetic field was slower than that for the $633 nm$ wavelength, and (b) 
for $780 nm$  (diode laser) for which too, the fall of intensity was
slower than that for $633 nm$ light. As we shall see below, these experimental
observations are all qualitatively consistent with what is expected from our 1D
model. }\\
\section{Discussion}
{We now consider more explicitly  how the applied magnetic field tunes the
scattering properties of the ferrofluid. There is ample
evidence that the nanoparticles (nanomagnets) rearrange to form spatially ordered
structures, like linear chains directed along the magnetic field \cite{Abraham, Butter, Li1, Li2}.  An
overall spatial ordering of the nanoparticles is to be expected on general grounds
as resulting from the minimization of the total energy of the magnetostatic coupling
of the nanomagnetic particles with  the externally applied magnetic field, and the
magnetic dipolar interaction between these nanomagnets. The permanent magnetic dipole moment of these single domain magnetite nanoparticles is ~$5 \times 10^{4}$ Bohr magneton. The magnetostatic energy per
nanoparticle in the external magnetic field of $300 G$ is about $0.1 eV$, exceeding both the thermal
energy ($k_BT$), and the inter-dipolar interaction magnetic energy
per nanoparticle. This implies that the external magnetic field can indeed introduce
spatial ordering of these nanomagnets. Thus, by varying the external magnetic field,
we can expect to continuously tune the spatial order (wave-vector $Q$), sweeping
across the Bragg resonance condition ($Q$ = 2$k_0$, \textit{i.e.}, $q = 2$), leading to  strong scattering.
The strong scattering (small mean free path) enhances the time taken by light to
transit through our LAB cavity. This should result in a dip in the transmission and the 
reflection coefficients bracketing the Bragg condition, as indeed our simulations (Fig. 1) show
and our experiments (Fig. 3) reaffirm. As noted before (following Eq. \ref{e.ReqT}) the asymmetric decrease of the two intensities arises from the simple fact that  light is incident from one end of the
sample, and is not injected directly  well within the sample (LAB). \\
Encouraged by this qualitative agreement between the numerical results of our
1D-model simulation and the experimental observations on our 3D ferrofluid, we now
proceed to examine if the choice of the parameter values for the model
(specifically, $\eta_{OR},  \eta_P,   \eta_I$)  do indeed correspond reasonably well
to our experimental ferrofluid. The input parameters taken for the ferrofluid in the
experiment are: nanoparticle mean number density $n_{FF} \simeq 10^{16}cm^{-3}$,
nanoparticle radius $a_{NP}\simeq 15$ nm, dielectric constant of the nanoparticle material
$Fe_3O_4, \epsilon_{NP} = 4 + i 2.2$, and the dielectric constant of the suspension medium
(kerosene) $\epsilon_K = 1.8$.  The above parameters straightforwardly give the mean
dielectric constant (the real part used in Eq. \ref{e.dRbydx} and \ref{e.dTbydx}, 
\begin{equation}
\label{e.epsilon0}
\epsilon_0 = \epsilon_K (1 -{{4 \pi}\over{3}} n_{FF} a_{NP}^3) + Re(\epsilon_{NP})
{{4 \pi}\over{3}}  a_{NP}^3n_{FF} \simeq 2.1
\end{equation}  
Next, the dissipation parameter :
\begin{equation}
\label{e.etaI} 
\eta_I = \epsilon_I / \epsilon_0 =
Im(\epsilon_{NP} ){\frac{4 \pi}{3}} n_{FF} a_{NP}^3 \simeq 0.31.
\end{equation}
 The random
 disorder parameter $\eta_{OR}$, arises out of the   fluctuations of the local number density of nanoparticles. Assuming a Poissonian distribution for the fluctuations and making use of the fact that the
variance equals the mean (both taken over the length scale of a wavelength), we obtain
\begin{equation}
\label{e.etaOR}
\eta_{OR}^2 = 32 \pi ^2 (Re \epsilon_F - \epsilon_k)^2({\frac{4 \pi} {3}} 
a_{NP}^3)^2 ({\frac{n_{FF}} {\lambda_0^3}}) \simeq 0.037 << 1 ,
\end{equation}
 as expected for a dense
random system (${{4 \pi} \over {3}}  \lambda^3_0 n_{FF} \gg
1$).  This again is consistent with the  fact that the  dense random system is not,
by  itself, an effective scatterer -- it is the  spatial modulation (Bragg
scattering) that makes the overall scattering effectively strong.} 

The  simultaneous minima of transmission and reflection at $q=2$ in
Fig.2 is a clear signature of the Bragg resonance condition, and is
consistent with the ferrofluid parameters. This  can be seen from the
following  geometrical considerations based on the fact that in a
magnetic field ($\sim$ 100 G) the nanoparticles are known to  re-arrange
forming  chains aligned parallel to the  field \cite {Abraham, Butter, Li1, Li2}. Assuming that the overall number
density of the ferrofluid remains  unchanged, we have  $1/n_{FF} = 2a_{NP}d^2$, where  $2a_{NP}$ 
 is the nanoparticle spacing along the chains and  $d$  the interchain separation. This gives $d \sim$half  the wavelength of light, $\lambda_0$ in the medium, and thus closely corresponds to the Bragg condition.  From the same geometrical considerations, the parameter $\eta_P$, giving the depth of the periodic dielectric modulation under the above Bragg condition, can be readily estimated.  It is given by $\eta_P = (16\pi/3) (\epsilon_{NP}/\epsilon_K) (n_{FF} a_{NP}^3) \simeq 1$.  Overall, given the complexity of the system and the simplicity of our model, we find the above estimates of the various parameters quite consistent with our  experiments.}\\

Next we turn to polarisation, namely, how, despite multiple scattering, the
transmitted as well as the reflected light retains the incident linear polarisation.
Let us  consider the case of transmission, where the initial and the final
wave-vectors are parallel.  For clarity, we resolve the
linearly polarised light into its left and  the right  circularly
polarised components, both of which traverse the same  path-length in
the optical  medium,   and, therefore, undergo the same \textit{dynamical} phase
shift. {This causes no change in  the state of polarisation. Phase shifts, however,
do also arise as a result of the \textit{geometric} phases suffered by the two circularly polarised 
components. These phase shifts are equal and opposite for the left and the right components, and have a magnitude given by
the solid angle subtended at the origin in the space of directions traced out
by the ray trajectory. For light, with photon spin =1, this does lead to a rotation of
the plane of linear polarisation by half that solid angle, which is, of course, random for our disordered system. {In
the present case, however, the scattering is dominated by the Bragg
scattering which is essentially in the forward and the backward directions. These multiple
scatterings subtend zero solid angle, and hence give no rotation (due to the geometric
phase)  of the plane of incident linear polarisation for the transmitted (forward)
and the reflected (backward) scattered light. Moreover,  recall that a single scattering by a Rayleigh scatterer (our spherical nanoparticle) retains the incident linear polarisation of light, anyway.  Besides, unlike the case in \cite{Li1,Li2}, our
magnetic field is transverse to the direction of incidence, and hence there is no Faraday
rotation. As already noted, in the transverse directions, the light intensities
are very small. }}\\
{Finally, our model can be viewed in a broader perspective: some of the  quantities
of possible physical interest become calculable. It is possible to calculate
the photon transit time ($\tau_L$) in our LAB cavity. Indeed, the introduction of a
weak dissipation ($\ell_i$) provides an effective measure \cite{Anantha} for $\tau_L$,
which can be readily shown to be given by  $e^{-\tau_L c_0 / \ell_i} =
(\left|T\right|^2 + \left|R\right|^2$). Also, the amount of light energy stored ($U_S$) in our
LAB cavity under steady-state illumination (intensity $I$) can then be readily
expressed as $U_S = I \tau_L$.  We note that this light is stored as {\em light},
and not as an electronic excitation. It is tempting to suggest here that this stored light may
re-emerge from the LAB cavity as a flash if the magnetic field (and, therefore, the confining
strong scattering) could be switched off instantaneously. This, of course, assumes that the nanoparticles in the ferrofluid relax back sufficiently fast. The emergent light pulse would 
have a linear polarisation same as that of  the incident light.   Clearly, this would not be the case if the light had been stored as an electronic excitation, and re-emitted as  fluorescence.  

\section{Summary}
In summary, we have studied experimentally as well as theoretically light transmission
through and its reflection from a magneto-statically tunable ferrofluid. We have
developed a mechanism in terms of scattering of light in a disordered medium in the
presence of weak dissipation, where the scattering could be effectively enhanced by a
spatial periodic modulation of the background dielectric constant, giving Bragg
scattering. Our theoretical treatment of transmission and reflection of light is based on the method of invariant imbedding.  The key idea to emerge from our studies is that the strong
multiple scattering of light and the resulting small mean free path can give rise to a long photon
sojourn time, thereby effectively making our ferrofluid system akin to a
lossy cavity. It is apt to point out here that the increased photon sojourn time calculated and reported in this work comes from  multiple scattering because of disorder, re-inforced by Bragg scattering due to spatial modulation. It does not arise because of dispersion  as in the well-known slow-wave systems, where  the slowness comes essentially from the fast variation of the real part of the refractive index with frequency, as, for example, in the  Electromagnetically Induced Transparency (EIT) systems, where very low group velocities (as low as few $cm s^{-1}$ could be achieved over a   narrow frequency window \cite{Hau}.  
Given the recent research activity \cite{Mehta, Kalpakkam, Li1, Li2} in the magneto-optical properties of magnetically tunable ferrofluids, the present work seems to be of some interest. \\

\section{Acknowledgement}
One of us (M. S.) wishes to acknowledge the Raman Research Institute for extensive use of  the experimental facilities and for hospitality during the course of this work.\\

{

}


\begin{thebibliography}{100}

\bibitem{Anderson}
  Anderson P. W.
  Phys. Rev. {\bf109} 1492 (1958).
  
\bibitem{Sheng}
Sheng P., 
Introduction to Wave Scattering, Localization and Mesoscopic Phenomena
 Academic Press, New York
  (1995).

\bibitem{Graham}
  see for example, Graham M., Adkins C. J., Behar H.  and Rosenblaum R.
  J. Phys. : Cond. Matter  {\bf10}  809 (1998).


\bibitem{JohnPhysToday}
John, S., Phys. Today, {\bf44}, 32 (1991).

\bibitem{JohnPRL}
 John, S., Phys. Rev. Lett.{\bf58} 2486 (1987). 

\bibitem{Kumar}
Kumar N., Phys. Rev. {\bf B31}  5513 (1985). 

\bibitem{Ramal}
Ramal R. and  Doucot B., J. d'Physique {\bf48} 509, (1987). 

\bibitem{Abraham}
Abraham V. S., Swapna Nair, Rajesh S., Sajeev U.S. and Anantharaman M. R.
 Bull. Mater. Sci. {\bf27} 155 (2004). 

\bibitem{Butter}
Butter K., Bomans P. H., Frederik P. M., Vroege G. J. and Philipse A. P. Nature Matl. {\bf2}, 88 (2003). 


\bibitem{Li1}
Li J., Liu X., Lin Y., Qui X., Ma X. and Huang Y.
 J. Phys.D: Appl. Phys. {\bf37}, 3357 (2004).

\bibitem{Li2}
Li J., Liu X., Lin Y., Bai L., Li Q. and Chen X.
Appl. Phys. Lett. {\bf91}, 253108 (2007).

\bibitem{Anantha}
Anantha Ramakrishna S. and Kumar N., 
Phys. Rev. {\bf B61}, 3163 (2000).


\bibitem{Hau}
Hau L.V., Harris S.E., Zachary Dutton and Cyrus Behroozi
Nature {\bf397} 594 (1999). 

\bibitem{Mehta}
Mehta R. V., Rajesh Patel, Rucha Desai, Upadhyay R. V. and Kinnari Parekh, Phys. Rev. Lett. {\bf 96}, 127402 (2006),  
see, however, Hema Ramachandran and Kumar N.
Phys. Rev. Lett. {\bf100} 229703 (2008).

\bibitem{Kalpakkam}
Laskar J. M., Philip J. and Baldev Raj
Phys. Rev. {\bf E78}, 031404 (2008). 
\end{thebibliography}
\end{document}